\documentclass[conference]{IEEEtran}
\usepackage{cite}
\usepackage{amsmath,amssymb,amsfonts}
\usepackage{graphicx}
\usepackage{textcomp}
\usepackage{booktabs}
\usepackage{multirow}
\usepackage{array}
\usepackage[T1]{fontenc}
\usepackage{xurl}
\graphicspath{{figures/}}
\begin{document}

\title{Decomposing Firm-Level Crisis Responses from Incomplete Market Signals: Evidence from China's IT Sector During COVID-19}

\author{\IEEEauthorblockN{Xiao Han}
\IEEEauthorblockA{Goizueta Business School\\
Emory University\\
Atlanta, Georgia, USA\\
xhan@alumni.emory.edu}
\and
\IEEEauthorblockN{Yao Xiao}
\IEEEauthorblockA{College of Computing\\
Georgia Institute of Technology\\
Atlanta, Georgia, USA\\
yxiao344@gatech.edu}}

\maketitle

\begin{abstract}
Exogenous shocks generate heterogeneous behavioral responses across firms, yet event studies typically report only sector-level averages. This paper develops a multi-method approach combining causal identification (difference-in-differences with cluster-robust inference), unsupervised behavioral discovery (K-means trajectory clustering, Gaussian hidden Markov models), and cross-sectional resilience prediction (logistic regression with nested cross-validation) to decompose firm-level response heterogeneity from noisy market signals. We demonstrate the approach on 246 Chinese A-share IT firms (216 with complete data for all analyses) during the COVID-19 shock (January 2020), using 252 non-IT CSI 300 firms as controls. The return decline was market-wide, not IT-specific (DID $p = 0.59$); the IT-specific effect was elevated volatility (DID $\beta = 0.043$, cluster-robust $p < 0.001$), with the effect surviving Benjamini-Hochberg correction in 13 of 30 alternative specifications. Unsupervised clustering produced three trajectory groups: fast recovery (36 companies, $+29.7\%$), resilient/moderate (67 companies), and persistent drag (113 companies, $-6.9\%$). Pre-crisis financial fundamentals showed only modest predictive power for resilience (nested CV AUC $= 0.635$, 95\% CI: 0.558--0.711; permutation $p = 0.016$), consistent with the limited informativeness of publicly available signals for anticipating crisis outcomes. The combination of causal analysis, unsupervised learning, and prediction represents a reproducible framework which can be applied to crises in other market periods.
\end{abstract}


\section{Introduction}

\subsection{The Problem: Decomposing Crisis Response}

Upon the occurrence of a shock affecting an industry, the traditional approach seeks to determine the average effect of the treatment. However, averages mask a great deal of heterogeneity, because some firms will recover in a matter of weeks, while others face a prolonged headwind; some firms become more volatile, while others stabilize. Separating those different effects from the mixed signals of the market is a common issue faced when analyzing crises, regardless of their origins, which could be anything from a pandemic to regulations.

Separating sectoral effects from market-wide changes is a causal inference problem; identifying latent structure in firm-level reactions without a priori labels is an unsupervised learning problem; and predicting resilience from pre-crisis observables, where the true answer may well be negative, is a prediction problem. These interrelated challenges call for an integrated multi-method approach.

\subsection{Empirical Setting: The COVID-19 Shock}

To demonstrate our proposed methodology, we consider the case study of how the IT industry in China responded to the news of the COVID-19 outbreak. This setting has several desirable properties: unambiguous shock timing (first reported fatality on January 11, 2020, a Saturday, with the first post-announcement trading day on January 13; Wuhan lockdown on January 23; WHO PHEIC declaration on January 30), genuine demand ambiguity within IT (work-from-home tailwinds for some firms, supply chain disruptions for others), and daily trading data for 246 firms over 242 trading days.

A distinctive feature of this event is the extended market closure: the Shanghai Stock Exchange was closed from January 24 through February 2 for the Lunar New Year holiday, which was extended by three days due to the outbreak. Trading resumed on February 3. This means that the market's first opportunity to react to the Wuhan lockdown (January 23, the last pre-closure trading day) and the WHO declaration (January 30, during the closure) was compressed into a single reopening day, making the shock window unusually concentrated.

\subsection{Research Questions and Contributions}

This study asks four questions about crisis response decomposition: (1)~Can sector-specific behavioral effects be isolated from market-wide shock responses? (2)~Do distinct trajectory groups emerge from firm-level post-shock signals? (3)~Do pre-crisis firm characteristics predict post-shock resilience? (4)~What is the temporal dimension of behavioral adaptation?

The main value of the paper lies in the methodology. While we do not assert a long-term pattern between COVID-19 and the Information Technology industry in China, we propose a multi-step methodology combining techniques for causal inference (difference-in-differences (DID) analysis with cluster-robust inference), unsupervised learning (using K-means and Gaussian hidden Markov models (HMMs)), and prediction (logistic regression with nested cross-validation) on the basis of an empirical study involving COVID-19. The proposed methodology provides a framework that, in theory, may be applied to other crises but has yet to be tested. The approach helps to understand the importance of weak or null results (e.g., the limited predictive value of fundamentals) in that it sets the limits of what can be extracted from noisy data.

The paper contributes in three ways. First, the COVID-19 impact on IT is transmitted via volatility rather than returns, an insight that stems from investor disagreement under demand uncertainty~\cite{bertrand2004,petersen2009}. Second, unsupervised trajectory analysis reveals that within-sector averages mask significant differences in how individual firms react. Third, prior public fundamentals have only modest predictive value for crisis resilience, with the model performing modestly above chance.

\section{Related Work}

Event studies provide the means to ascertain abnormal returns around shocks~\cite{mackinlay1997}. Estimation based on the DID method with clustered standard errors~\cite{bertrand2004,petersen2009} and event-study specifications~\cite{sun2021} form a standard approach in panel identification. Previous studies of shocks have typically been limited to sector- or market-level averages: Yang and Lv~\cite{yang2014} detected temporary impacts from isolated events while Baker et al.~\cite{baker2020} demonstrated that COVID-19 brought more lasting volatility than SARS~\cite{david2021}; Liu~\cite{liu2018} recorded adverse shocks on Asian equities at the time of financial crises. Hidden Markov models, introduced to finance by Hamilton~\cite{hamilton1989}, facilitate detection of latent regimes in the market, while clustering contributes to differentiation of companies according to their response curves; however, few works integrate causal identification with unsupervised decomposition of individual events. Models of heterogeneity of investors~\cite{hong1999,delong1990} show that in crises the diversity of opinion rises~\cite{barberis2003}, resulting in price fluctuations being detached from fundamentals~\cite{shiller2000}. In particular, studies of COVID-19 reveal the following: Wu et al.~\cite{wu2020} recorded staged shocks in the Chinese equity market; He et al.~\cite{he2020} established different responses for different sectors; Ding et al.~\cite{ding2021,ding2020b} showed that digital maturity affected pandemic impacts; Zhang et al.~\cite{zhang2020} found that contagion effects increased with investor sentiment dominating fundamentals; and Chen et al.~\cite{chen2022} demonstrate that fear sentiment and policy uncertainty drove asset price dynamics. Overall, however, the existing studies mainly focus on sector-level averages~\cite{he2020} or cross-country comparisons~\cite{liu2020,phan2020} and do not take into account within-sector heterogeneity.

\section{Data and Methodology}

\subsection{Sample Selection and Reconciliation}

We study A-share listed companies in China's IT sector under the CSRC 2012 taxonomy (``Information Transmission, Software, and Information Technology Services''), comprising 230 software/IT services firms and 16 telecommunications firms. The reference event date is January~11, 2020 (first officially reported COVID-19 fatality). Since January~11 fell on a Saturday, the first post-announcement trading day is January~13, 2020. The panel spans June~3, 2019 to May~29, 2020 (242 unique trading days), restricted to 2020~H1 to avoid confounding from subsequent events.

\textit{Sample flow.} As shown in Table~\ref{tab:sample}, the sample is tracked from identification through each stage of analysis. The panel is unbalanced, meaning that not all 246 firms are observed on all 242 trading dates. The 54,040 firm-day observations represent 90.8\% of the theoretical maximum ($246 \times 242 = 59{,}532$); 41 firms have fewer than 200 trading days, mainly due to recent listing dates or trading suspensions. Missing days are not imputed; each analysis uses only the available observations for each firm.

\begin{table}[htbp]
\caption{Sample Reconciliation}
\begin{center}
\footnotesize
\begin{tabular}{lcc}
\toprule
\textbf{Stage} & \textbf{IT Firms} & \textbf{Reason for Reduction} \\
\midrule
Identified (CSRC) & 246 & -- \\
Q4 2019 fundamentals & 223 & 23 lack financial reports \\
DID treatment group & 218 & 28 lack sufficient pre-period data \\
Resilience model & 216 & 30 lack complete predictors \\
\bottomrule
\end{tabular}
\label{tab:sample}
\end{center}
\end{table}

\textit{Control group.} In the case of DID analysis, 252 non-IT CSI 300 companies serve as controls. These companies are included in the broad market index and characterize large-cap stocks across all non-IT sectors. A Mahalanobis-matched subsample (218 pairs, matched on pre-period mean turnover and volatility) addresses the size differentials between large-cap CSI 300 controls and predominantly small/mid-cap IT firms. Pre-period balance is reported in Table~\ref{tab:balance}. The standardized mean differences exceed the conventional threshold of $|SMD| < 0.1$ for all covariates, indicating that IT and non-IT firms differ substantially in baseline market characteristics. The matching was successful in reducing these differences (matched-sample volatility DID $\beta = 0.036$, $p < 0.001$), but residual imbalance should be kept in mind when interpreting the DID estimates.

\begin{table}[htbp]
\caption{Pre-Period Covariate Balance: IT vs. Control}
\begin{center}
\footnotesize
\begin{tabular}{lcccc}
\toprule
\textbf{Variable} & \textbf{IT Mean} & \textbf{Control Mean} & \textbf{SMD} & $p$ \\
\midrule
Mean Return & 0.0018 & 0.0010 & 0.42 & $<$0.001 \\
Mean Volatility & 0.418 & 0.271 & 1.57 & $<$0.001 \\
Mean Turnover & 0.033 & 0.010 & 1.05 & $<$0.001 \\
Return Std. & 0.027 & 0.018 & 1.55 & $<$0.001 \\
\bottomrule
\end{tabular}
\label{tab:balance}
\end{center}
\end{table}

\subsection{Variable Definitions}

\textit{Dependent variables:} Daily stock return ($Yield_{it}$) and daily return volatility ($Volatility_{it}$), defined as the annualized standard deviation of daily log returns over a rolling 20-trading-day window. Both IT and control groups use identical computation to ensure comparability.

\textit{Rolling window considerations.} The rolling 20-day volatility analysis involves two considerations. First, the use of overlapping rolling windows leads to serial dependence in the volatility series, for which we cluster standard errors at the firm level. Second, in the first 19 post-event observations of volatility, we mechanically include returns prior to the event. We address this in three ways: (a)~excluding the first 20 post-event trading days, (b)~using daily absolute and squared returns as non-overlapping proxies, and (c)~computing non-overlapping 20-day block volatility (see Section~\ref{sec:vol_robust}).

\textit{Independent variable:} $Post_t = 1$ for trading days from January 13, 2020 (the first post-announcement trading day) through May 29, 2020. \textit{Controls:} Stock turnover rate ($Turnover_{it}$). The analysis relies on locally archived Wind database extracts (see Limitations).

Table~\ref{tab:descriptive} reports descriptive statistics for the IT panel of 54,040 firm-day observations.

\begin{table}[htbp]
\caption{Descriptive Statistics ($N = 54{,}040$ firm-days)}
\begin{center}
\begin{tabular}{lccccc}
\toprule
\textbf{Variable} & \textbf{Mean} & \textbf{Std.} & \textbf{Min} & \textbf{Max} & \textbf{Median} \\
\midrule
Yield & 0.0008 & 0.0324 & $-$0.102 & 0.102 & 0.000 \\
Volatility & 0.4823 & 0.1855 & 0.093 & 1.312 & 0.453 \\
Turnover & 0.0367 & 0.0404 & 0.0001 & 0.603 & 0.024 \\
\bottomrule
\end{tabular}
\label{tab:descriptive}
\end{center}
\end{table}

\subsection{Methods}

We apply interrupted time-series analysis~\cite{cook1979} and two-way fixed-effects panel models with cluster-robust standard errors. The DID specification identifies IT-specific effects:

\begin{equation}
Y_{it} = \beta_3 \cdot (IT_i \times Post_t) + \mu_i + \lambda_t + \varepsilon_{it}
\label{eq:did}
\end{equation}

\noindent where $\mu_i$ are firm fixed effects, $\lambda_t$ are time fixed effects, and standard errors are clustered by firm. The main effects $IT_i$ and $Post_t$ are absorbed by firm and time fixed effects respectively and are omitted from the equation. Turnover is excluded from the DID specification because it is potentially endogenous to treatment: the pandemic shock itself may alter trading behavior differentially for IT versus non-IT firms, making it a post-treatment variable~\cite{angrist2009}.

To assess the parallel trends assumption, we estimate an event-study specification replacing the single $IT_i \times Post_t$ interaction with a full set of week-by-treatment interactions, omitting the week immediately before the event as the reference period. This produces lead and lag coefficients with cluster-robust confidence intervals.

K-means clustering~\cite{hartigan1979} on standardized trajectory features discovers response groups; we report results for $k = 2, 3, 4$ and assess stability across 50 random initializations using the adjusted Rand index (ARI). Logistic regression with nested cross-validation (5-fold outer, 3-fold inner for hyperparameter selection) and bootstrap confidence intervals predicts resilience from 13 pre-crisis variables. Gaussian HMMs~\cite{rabiner1989} model latent market states; we report BIC and AIC for 2 through 5 states to justify model selection.

\section{Results}

\subsection{Panel Estimates}

Interrupted time-series (ITS) analysis confirms a structural break at announcement: returns drop ($\beta = -0.006$, $t = -7.55$, $p < 0.01$) and volatility spikes ($\beta = 0.003$, $t = 4.46$, $p < 0.01$). A Hausman test ($p = 0.9997$) supports the random-effects specification for ITS, though we note this test may be unreliable in the presence of cluster-robust standard errors; fixed-effects ITS results are qualitatively identical.

\begin{figure*}[htbp]
\centerline{\includegraphics[width=0.9\textwidth]{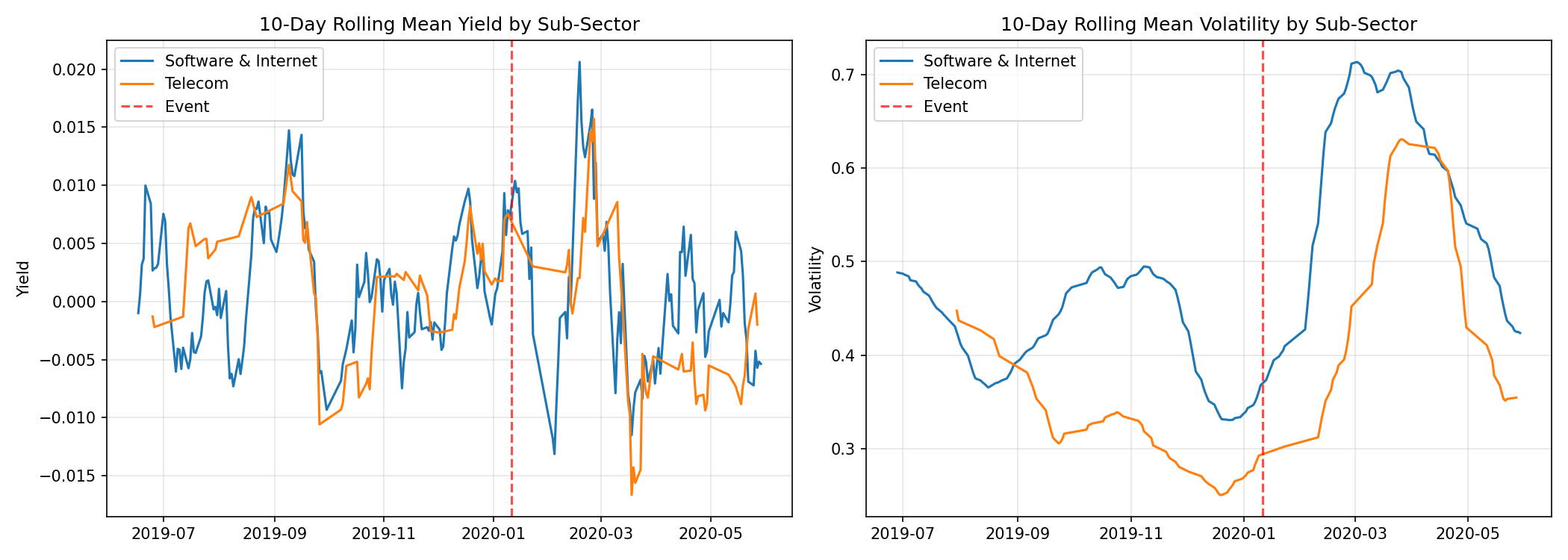}}
\caption{Structural break identification: the IT sector's returns and volatility show a clear discontinuity around the COVID-19 announcement date. The vertical line marks January 13, 2020, the first trading day after the first reported fatality. The market was closed January 24 through February 2 for the extended Lunar New Year holiday.}
\label{fig:sector}
\end{figure*}

Table~\ref{tab:fe} presents IT-only panel regressions. Columns (1) and (3) include firm fixed effects only, with the $Post$ dummy identified directly. Columns (2) and (4) add date fixed effects. In this specification, the $Post$ dummy is nearly collinear with the date indicators; the coefficient is identified only because the panel is unbalanced (9.2\% of firm-day cells are missing), so it captures residual variation across the subset of firms observed on each date rather than a clean pre/post contrast. We therefore treat columns (2) and (4) as specification checks rather than primary estimates. The IT-specific effect is identified more cleanly by the DID design (Section~\ref{sec:did}), where the interaction term varies across both firms and time.

\begin{table}[htbp]
\caption{IT-Only Panel Regression (Cluster-Robust SE)}
\begin{center}
\begin{tabular}{lcccc}
\toprule
& \multicolumn{2}{c}{\textbf{Yield}} & \multicolumn{2}{c}{\textbf{Volatility}} \\
\cmidrule(lr){2-3} \cmidrule(lr){4-5}
& (1) & (2) & (3) & (4) \\
\midrule
Post & $-$0.0018*** & $-$0.0008 & 0.1553*** & 0.0796*** \\
& (0.0003) & (0.0036) & (0.0014) & (0.0183) \\
Firm FE & Yes & Yes & Yes & Yes \\
Date FE & No & Yes$^\dagger$ & No & Yes$^\dagger$ \\
\bottomrule
\multicolumn{5}{l}{\footnotesize *** $p<0.01$. $^\dagger$Post coefficient near-collinear with date FE;} \\
\multicolumn{5}{l}{\footnotesize identified only from panel imbalance (see text).} \\
\end{tabular}
\label{tab:fe}
\end{center}
\end{table}

Sub-sector analysis reveals heterogeneity: software/internet firms show significant return declines and volatility increases, while telecom shows larger return declines but \textit{negative} volatility ($-0.0087$, $p < 0.05$), consistent with inelastic demand. The telecom sub-sample ($N = 16$) should be interpreted qualitatively.

\subsection{Robustness}
\label{sec:robust}

\textbf{Placebo tests} were conducted before the event of interest, using pseudo-event windows in non-COVID periods (1, 2, and 3 months before the actual event). We observe coefficients with reversed signs (negative for volatility) and magnitudes ranging from 0.01 to 0.04, with standard errors mostly overlapping zero. Even if the sign reversal is in accordance with the effect being event-specific, the placebo estimates are imprecise and do not by themselves rule out all confounders.

\textbf{Alternative event dates and windows:} Three event dates (January 11, January 23, January 30) are evaluated through five windows (30, 60, 90, 120 trading days, and the full sample) for two outcomes, totaling 30 specifications ($3 \times 5 \times 2$) with Benjamini-Hochberg FDR correction at $q = 0.05$. Only 13 of 30 specifications are statistically significant after correction. All significant full-window and 60-day-window specifications are on the volatility outcome, and all three event dates in the full-sample window yield statistically significant effects (coefficients 0.078--0.080, $p < 0.01$). However, results based on the 30-day window should be treated with caution, as the limited duration of the window and the market closure lead to inflated coefficients (up to $\beta = 0.58$ for volatility) that may not correspond to sustained effects.

\textbf{Phase decomposition:} Replacing the binary Post dummy with four phase indicators shows the return decline clusters within the lockdown period (January 23--30). Elevated volatility during Recovery (March onward) likely reflects the global market crash.

\textbf{Synthetic control:} A synthetic IT-sector counterfactual from 252 non-IT firms~\cite{abadie2010} shows IT firms outperformed cumulatively by $+8.57\%$, consistent with the panel finding that IT experienced a larger initial decline but faster recovery.

\subsubsection{Volatility Measure Robustness}
\label{sec:vol_robust}

Four alternative volatility specifications are examined to address concerns related to the rolling window: (1)~daily absolute returns ($p = 0.14$), (2)~daily squared returns ($p = 0.37$), (3)~dropping the first 20 post-event days ($\beta = 0.070$, $p = 0.018$), and (4)~non-overlapping 20-day blocks ($p = 0.42$). Specification~3 yields a significant result even after the contaminated observations are removed, while the other specifications do not produce significant results, which indicates that there is a regime shift that can be detected through multi-day aggregation but not in single-day returns.

\subsection{Difference-in-Differences: IT vs.\ Non-IT}
\label{sec:did}

Table~\ref{tab:did} reports DID estimates with both conventional and cluster-robust standard errors. The return coefficient is insignificant, confirming that the decline is market-wide. The volatility coefficient is significant: IT firms show 0.043 units higher volatility (cluster-robust $p < 0.001$). The 5.7$\times$ SE inflation from clustering is consistent with Bertrand et al.~\cite{bertrand2004}.

\begin{table}[htbp]
\caption{DID Estimates: IT vs.\ Non-IT CSI 300}
\begin{center}
\footnotesize
\begin{tabular}{lcc}
\toprule
& \textbf{Return} & \textbf{Volatility} \\
\midrule
$\beta_3$ (IT $\times$ Post) & $-$0.000135 & 0.0433 \\
\midrule
Conv.\ SE & 0.00031 & 0.00177 \\
$t$ / $p$ & $-$0.44 / 0.66 & 24.45 / $<$0.001 \\
Cluster-robust SE & 0.00025 & 0.01013 \\
$t$ / $p$ & $-$0.54 / 0.59 & 4.27 / $<$0.001 \\
95\% CI (cluster) & [$-$0.00063, 0.00036] & [0.0234, 0.0631] \\
\midrule
\multicolumn{3}{l}{$N$: 218 IT (treatment), 252 non-IT (control).} \\
\multicolumn{3}{l}{Mahalanobis-matched subsample (218 pairs): Vol $\beta = 0.036$, $p < 0.001$.} \\
\bottomrule
\end{tabular}
\label{tab:did}
\end{center}
\end{table}

\begin{figure}[htbp]
\centerline{\includegraphics[width=\columnwidth]{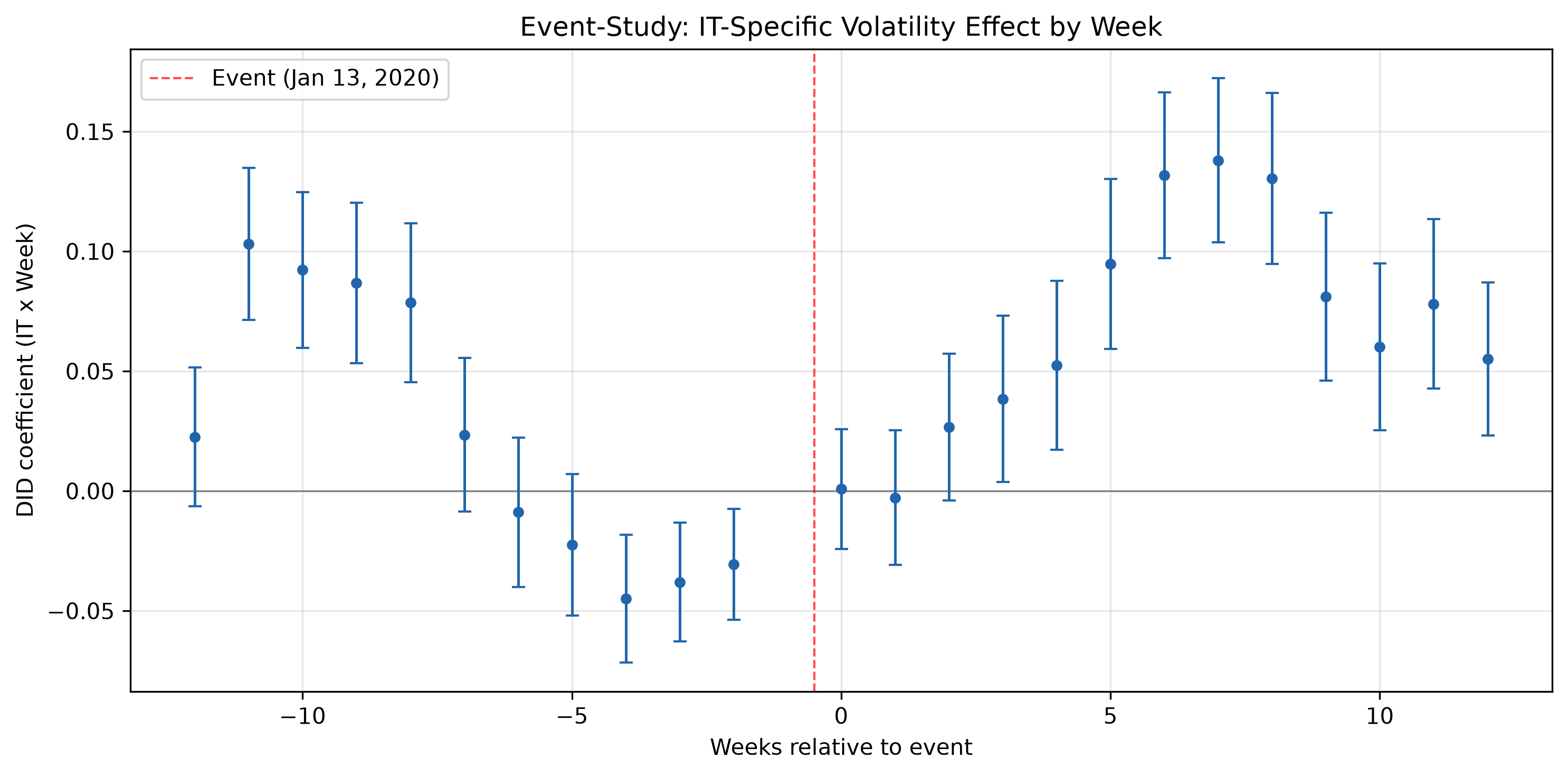}}
\caption{Event-study: weekly DID coefficients for IT-specific volatility, with 95\% cluster-robust confidence intervals. Reference period is $w = -1$. Pre-event coefficients are not uniformly zero, indicating that parallel trends are not cleanly met.}
\label{fig:event_study}
\end{figure}

\textbf{Event-study specification.} The coefficients for volatility presented in Figure~\ref{fig:event_study} (reference: $w = -1$) indicate that there are differences in volatility during the pre-event and post-event periods. The pre-event coefficient values do not all equal zero, as there is positive and significant IT-specific volatility in weeks $-11$ through $-8$, while weeks $-4$ through $-2$ exhibit negative and significant values. This shows that parallel trends are not cleanly satisfied. The post-event coefficients increase from around week~3 to weeks 6--7 ($\beta \approx 0.13$), indicating that the market crash that occurred in March 2020 is the reason for this increase. The conclusion that the post-event increase is simply a continuation of the pre-event trend can be countered by two features: first, the non-monotonicity in the pre-event pattern, and second, that the matched-sample DID ($\beta = 0.036$, $p < 0.001$) produces qualitatively similar results. Nevertheless, the DID volatility estimate should be read as suggestive rather than cleanly causal.

The IT-specific volatility premium reflects unresolved uncertainty surrounding competing demand shifts within IT~\cite{bloom2009}, amplified by increased investor disagreement~\cite{barberis2003}.

\subsection{Firm-Level Resilience Prediction}

We predict resilience (above-median cumulative return over 60 trading days from January 13) from 13 pre-crisis variables: 4 market variables, 1 size proxy, and 8 fundamentals from Q4 2019 reports ($N = 216$, positive class 47.7\%).

The feature scaling technique is applied in every CV fold so as to avoid data leakage. The outer and inner CVs are arranged in a nested manner for the purpose of hyperparameter selection (5-fold outer, 3-fold inner, $C \in \{0.01, 0.1, 1.0, 10.0\}$). The nested CV AUC has been found to be $0.635$ (95\% CI: 0.558--0.711; L2: $0.610 \pm 0.027$; L1: $0.632 \pm 0.030$), while the Brier score was computed to be 0.244---better than the naive model that had a Brier score of 0.250. Although CIs from bootstrapping do cover zero, the results of the permutation test ($n = 1{,}000$) yielded $p = 0.016$, implying that the joint model performs better than chance; however, no single feature is reliable on its own. Overall, it can be concluded that pre-crisis public fundamentals have some value in predicting post-shock resilience, although not enough for practical firm-level prediction. Omitted variables (R\&D intensity, institutional ownership) could potentially improve results.

\subsection{Latent States and Trajectory Groups}

\textbf{HMM model selection.} Gaussian HMMs~\cite{rabiner1989} are used to analyze the aggregate IT sector across 2--5 states. The Bayesian Information Criterion (BIC) favors a model with four states (BIC $= 1{,}396$) over three states (BIC $= 1{,}463$); however, the four-state model subdivides the Crisis state into two substates with fewer than 10 days each. For the sake of interpretability ($\Delta$BIC $= 67$ is modest relative to the added complexity), we retain a three-state model, treating HMM results as exploratory. The three-state model identifies Crisis (92 days), Normal (55 days), and Recovery (86 days), characterized by high self-transition probabilities (0.953--0.982). Per-firm HMM validates the K-means groups: persistent-drag firms take 3.4 days to exit Crisis versus 0.8 for fast-recovery firms.

\textbf{K-means model selection and stability.} The results in Table~\ref{tab:kmeans_stability} indicate both silhouette scores and cross-seed stability (using adjusted Rand index across 50 different initializations) for the values of $k = 2, 3, 4$. The silhouette score is highest for $k = 2$ (0.296), while $k = 3$ gives a moderately lower score of 0.237. The stability for $k = 3$ is found to be highly consistent across seeds (mean ARI $= 0.952$, minimum 0.847); however, $k = 4$ has considerably lower stability (mean ARI $= 0.712$, minimum 0.381).

\begin{table}[htbp]
\caption{K-Means Model Selection and Seed Stability}
\begin{center}
\begin{tabular}{lccc}
\toprule
$k$ & \textbf{Silhouette} & \textbf{Mean ARI} & \textbf{ARI Range} \\
\midrule
2 & $0.296 \pm 0.003$ & $0.962 \pm 0.035$ & [0.925, 1.000] \\
3 & $0.237 \pm 0.004$ & $0.952 \pm 0.044$ & [0.847, 1.000] \\
4 & $0.215 \pm 0.005$ & $0.712 \pm 0.156$ & [0.381, 1.000] \\
\bottomrule
\end{tabular}
\label{tab:kmeans_stability}
\end{center}
\end{table}

We select $k = 3$ because $k = 2$ merges fast-recovery ($+29.7\%$) and resilient/moderate ($-1.0\%$) firms, obscuring economically meaningful distinctions. The high ARI ($0.952$) confirms reproducibility. K-means clustering on seven standardized trajectory features (cumulative return at 10, 30, and 60 days; days to recovery; volatility change; maximum drawdown; maximum rebound; $N = 216$) produces three clusters (Table~\ref{tab:clusters}), with stable group medians of 38, 64, and 113 firms across seeds.

\begin{table}[htbp]
\caption{K-Means Trajectory Groups ($k = 3$, silhouette $= 0.24$)}
\begin{center}
\begin{tabular}{lcccc}
\toprule
\textbf{Group} & $N$ & \textbf{60d Ret.} & \textbf{Max DD} & \textbf{Composition} \\
\midrule
Fast-Recovery & 36 & $+$29.7\% & Moderate & 35 SW/Int + 1 Tel \\
Resilient/Moderate & 67 & $-$1.0\% & Smallest & 14 Tel + 53 SW \\
Persistent-Drag & 113 & $-$6.9\% & Largest & All SW/Int \\
\bottomrule
\end{tabular}
\label{tab:clusters}
\end{center}
\end{table}

The group-sector relationship is significant ($\chi^2 = 29.6$, $p < 0.001$) but partly mechanical: 14 of 15 telecom firms fall into Resilient/Moderate, unsurprising given $N = 15$. Software/internet firms span all three groups.

\begin{figure}[htbp]
\centerline{\includegraphics[width=\columnwidth]{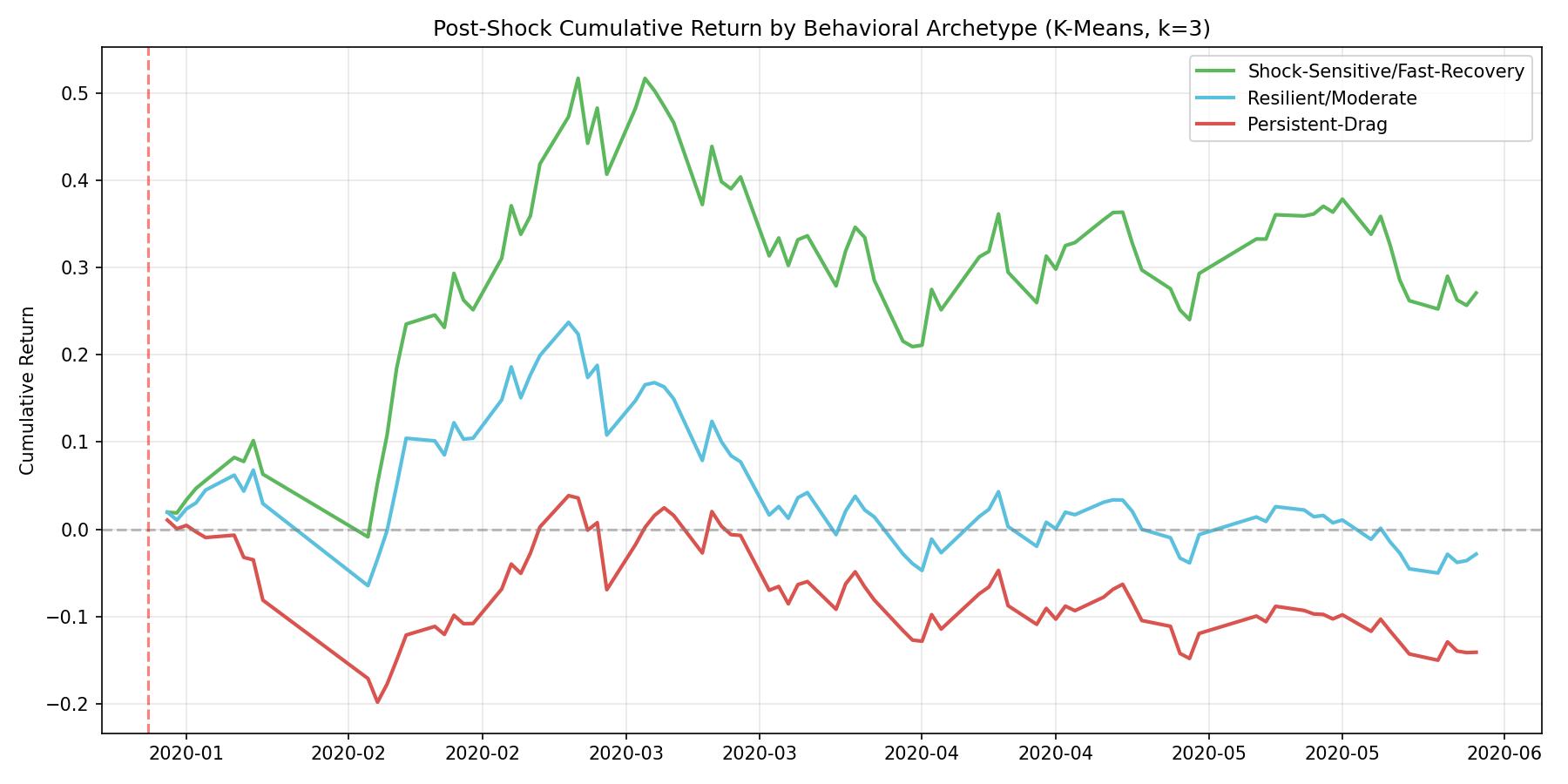}}
\caption{Post-shock trajectory by group. Fast-recovery firms achieved a mean 60-day cumulative return of $+29.7\%$; persistent-drag firms averaged $-6.9\%$, a gap of 36.6 percentage points. Resilient/moderate firms remained near baseline ($-1.0\%$).}
\label{fig:trajectories}
\end{figure}

\section{Conclusion}

The multi-method pipeline applied here to examine the impact of COVID-19 on China's IT sector comes to three findings. Firstly, the IT-specific effect is largely reflected in volatility rather than in returns (DID $\beta = 0.043$, cluster-robust $p < 0.001$, where results remain statistically significant after BH correction in 13 of 30 specifications), despite the event study indicating the presence of certain violations of the parallel trends assumption. Secondly, the use of unsupervised clustering methods produces three stable trajectory groups (ARI $= 0.952$), concealing a 36.6 percentage-point gap between fast-recovery ($+29.7\%$) and persistent-drag ($-6.9\%$) firms. Finally, the predictive power of the pre-crisis data is rather limited (nested CV AUC $= 0.635$; permutation $p = 0.016$), even though no single factor shows sufficient predictive capability.

The volatility channel indicates that simply focusing on price levels is not enough for regulators. From investors' perspective, the 36.6-percentage-point gap could not be predicted from pre-crisis public data, and the heterogeneous recovery contradicts the narrative that IT broadly benefited from COVID. This framework could be applied to other sectors affected by exogenous shocks.

\textit{Limitations.} The application of the DID approach to identify the treatment effect encounters challenges: the parallel trend assumption is violated at the pre-treatment stage, and substantial covariate imbalance (SMD $> 1.0$ for volatility, turnover, and return standard deviation) remains despite Mahalanobis matching; future work should consider the CSI IT sub-index as a better-matched control, and could incorporate sequence-based~\cite{xu2025mamba} and spatiotemporal~\cite{liang2026spatiotemporal} risk models. The rolling 20-day volatility measure creates serial dependence; the effect survives when contaminated observations are excluded ($p = 0.018$), however the daily proxies do not support the result, indicating dependence on multi-day aggregation. In applying the unsupervised methods, emphasis was put not on the best statistical performance but rather on interpretability (silhouette favors $k = 2$; BIC favors four-state HMM), with stability analysis confirming reproducibility. The resilience model's modest AUC may partly reflect data limitations: potentially relevant variables (R\&D intensity, institutional holdings) were unavailable after university database access ended. Finally, this is a single crisis, single market, single sector study that examines how investors behave in crises, not how firms operate. Analysis code is available at \url{https://github.com/XH8851/Decomposing-Firm-Level-Crisis-Responses-from-Incomplete-Market-Signals}; raw data are sourced from the Wind Financial Terminal under a university license.

\bibliographystyle{IEEEtran}

\end{document}